\documentstyle[graphicx,epsfig,12pt]{article}
\title{{\LARGE\bf Laser cooling of electron beams at linear
  colliders}.\thanks{Invited talk at the
  International Symposium on New Visions in Laser-Beam Interactions,
  October 11-15, 1999, Tokyo, Metropolitan University Tokyo, Japan. To
  be published in Nucl. Instr. and Meth. B.}  }

\author{Valery Telnov, \\
  {\small\it Budker Institute of Nuclear Physics, 630090 Novosibirsk, 
  Russia}\thanks{email:telnov@inp.nsk.su} \\
  {\small\it and DESY, Germany}} 

\date{}
\begin{document}
\newcommand{\EP}{\mbox{e$^+$}}
\newcommand{\EM}{\mbox{e$^-$}}
\newcommand{\EPEM}{\mbox{e$^+$e$^-$}}
\newcommand{\EMEM}{\mbox{e$^-$e$^-$}}
\newcommand{\EE}{\mbox{ee}}
\newcommand{\GG}{\mbox{$\gamma\gamma$}}
\newcommand{\GP}{\mbox{$\gamma$e$^+$}}
\newcommand{\GE}{\mbox{$\gamma$e}}
\newcommand{\LGE}{\mbox{$L_{\GE}$}}
\newcommand{\LGG}{\mbox{$L_{\GG}$}}
\newcommand{\LEE}{\mbox{$L_{\EE}$}}
\newcommand{\TEV}{\mbox{TeV}}
\newcommand{\GEV}{\mbox{GeV}}
\newcommand{\EV}{\mbox{eV}}
\newcommand{\CM}{\mbox{cm}}
\newcommand{\M}{\mbox{m}}
\newcommand{\MM}{\mbox{mm}}
\newcommand{\NM}{\mbox{nm}}
\newcommand{\MKM}{\mbox{$\mu$m}}
\newcommand{\E}{\mbox{$\epsilon$}}
\newcommand{\EN}{\mbox{$\epsilon_n$}}
\newcommand{\EI}{\mbox{$\epsilon_i$}}
\newcommand{\ENI}{\mbox{$\epsilon_{ni}$}}
\newcommand{\ENX}{\mbox{$\epsilon_{nx}$}}
\newcommand{\ENY}{\mbox{$\epsilon_{ny}$}}
\newcommand{\EX}{\mbox{$\epsilon_x$}}
\newcommand{\EY}{\mbox{$\epsilon_y$}}
\newcommand{\SEC}{\mbox{s}}
\newcommand{\CMS}{\mbox{cm$^{-2}$s$^{-1}$}}
\newcommand{\MRAD}{\mbox{mrad}}
\newcommand{\IND}{\hspace*{\parindent}}
\newcommand{\beq}{\begin{equation}}
\newcommand{\eeq}{\end{equation}}
\newcommand{\beqn}{\begin{eqnarray}}
\newcommand{\eeqn}{\end{eqnarray}}
\newcommand{\dst}{\displaystyle}
\newcommand{\bm}{\boldmath}
\newcommand{\BX}{\mbox{$\beta_x$}}
\newcommand{\BY}{\mbox{$\beta_y$}}
\newcommand{\BI}{\mbox{$\beta_i$}}
\newcommand{\SX}{\mbox{$\sigma_x$}}
\newcommand{\SY}{\mbox{$\sigma_y$}}
\newcommand{\SZ}{\mbox{$\sigma_z$}}
\newcommand{\SI}{\mbox{$\sigma_i$}}
\newcommand{\SIP}{\mbox{$\sigma_i^{\prime}$}}
\newcommand{\n}{\mbox{$n_f$}}
\maketitle
\begin{abstract}
 
  A method of electron beam cooling is considered which can be used
  for linear colliders. The electron beam is cooled during collision
  with focused powerful laser pulse. The ultimate transverse
  emittances are much below those achievable by other methods.  This
  method is especially useful for  high energy gamma-gamma colliders.
  In this paper we review and analyse limitations in this method, also
  discuss a new method of obtaining very high laser powers required for
  the laser cooling, radiation conditions and finaly present a possible scheme
for the laser cooling of electron beams.

\end{abstract}

\section{Introduction, one pass laser cooling}

To explore the energy region beyond LEP-II, linear colliders (LC) with
center--of--mass energy 0.5--2 TeV are developed now in the main
accelerator centers \cite{NLC},\cite{TESLA},\cite{JLC}. Besides \EPEM\ 
collisions, at linear colliders one can ``convert'' electrons to high energy
photons using the Compton backscattering of laser light, thus
obtaining \GG\ and $\gamma e$ collisions with energies and
luminosities close to those in \EPEM\ collisions
\cite{GKST81}-\cite{Tsit1}.

To attain high luminosity, beams in linear colliders should be very
tiny. At the interaction point (IP) in the current LC designs, beams
with transverse sizes as low as \SX/\SY\ $\sim$ 300/3 nm are planned.
Beams for \EPEM\ collisions should be flat in order to reduce
beamstrahlung energy loss. For \GG\ collision, the beamstrahlung
radiation is absent also there are no beam instabilities therefore
beams with smaller \SX\ ($\sigma_x/\sigma_y \sim 10/1$ nm) can be used
\cite{TSB2},\cite{TKEK}-\cite{Tsit1} to obtain higher luminosity.

The transverse beam sizes are determined by the emittances \EX, and \EY.
The beam sizes at the interaction point (IP) are $\sigma_i=\sqrt{\EI\
\BI\ }$, where \BI\ is the beta function at the IP.  With the
increase of the beam energy  the emittance of the bunch
decreases: $\EI=\ENI/\gamma$, where $\gamma=E/mc^2, $ \ENI\ is the
{\it normalized} emittance.

The beams with a small \ENI\ are usually prepared in damping rings
which naturally produce bunches with $\ENY\ll\ENX$ \cite{WID}.  Laser
RF photoguns can also produce beams with low emittances \cite{TRAV}.
However, for linear colliders it is desirable to have smaller emittances.

Recently, a new method of electron beam cooling was proposed which
allows further reduction of the transverse emittances after damping
rings or guns by 1--3 orders of magnitude \cite{TSB1},\cite{Monter}.

\begin{figure}[!htb]
\centering
\vspace*{-0.0cm} 
\hspace*{-0.4cm} \epsfig{file=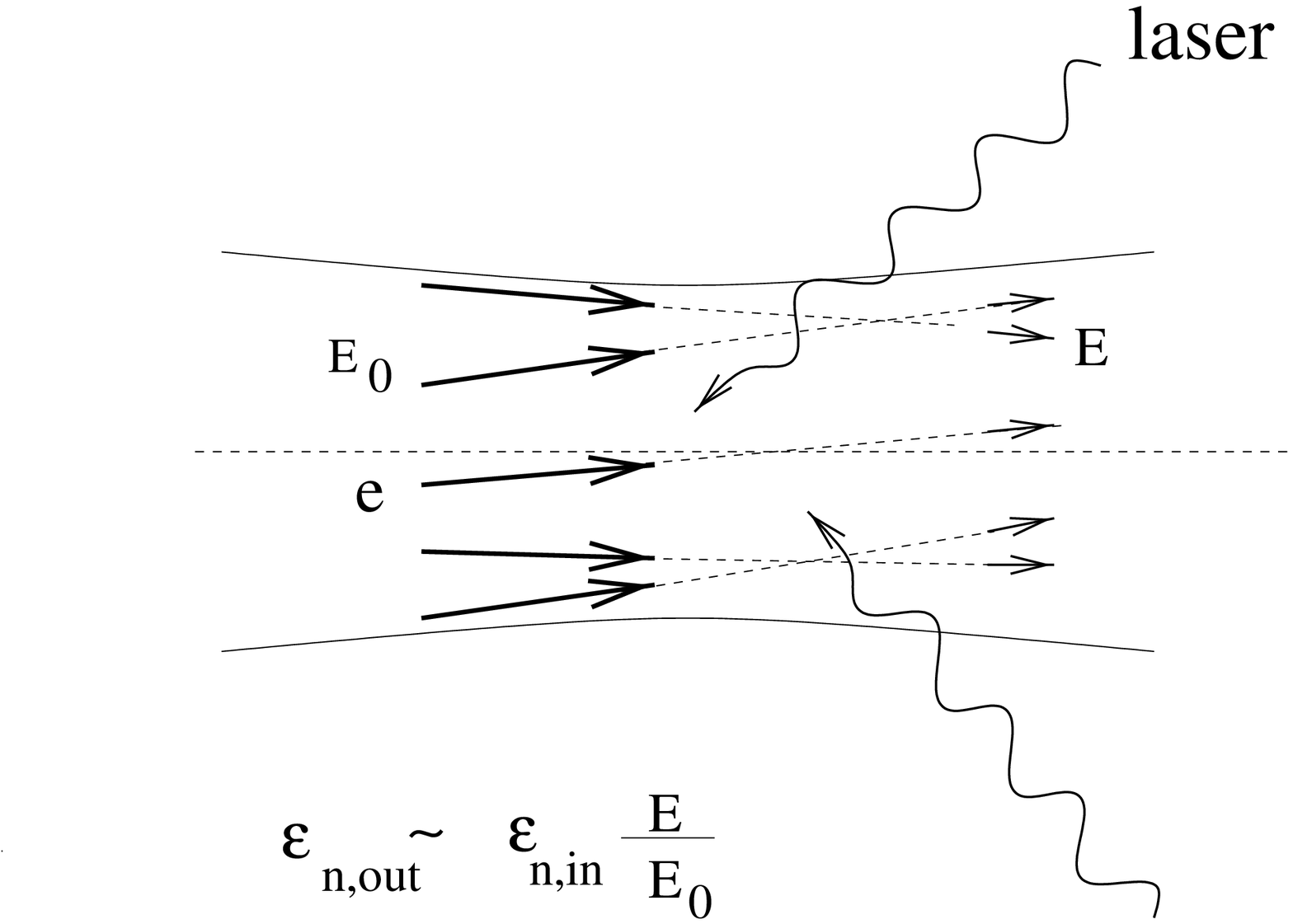,width=10cm,angle=0} 
\vspace*{-0.0cm} 
\caption{Principle of the laser cooling.}
\vspace{0mm}
\label{lcool}
\vspace{4mm}
\end{figure}

The idea of laser cooling of electron beams is very simple, \footnote{This idea was
  mentioned in the talk given by B.Palmer at the Berkeley Workshop on
  Gamma--Gamma colliders \cite{PALMER}, first analyses of this method was
  done in ref.\cite{TSB1}} see fig.~\ref{lcool}. 
 During head-on
collisions with optical laser photons (with an electromagnetic wave in
the case of a very strong field) the transverse distribution of electrons
($\sigma_i$) remains almost unchanged. Also the angular spread
($\sigma_i^{\prime}$) is almost constant, because for photon energies
(a few eV) much lower than the electron beam energy (several GeV) the
scattered photons follow the initial electron trajectory with a small
additional spread. So, the emittance $\EI = \SI \SIP$ remains almost
unchanged. At the same time, the electron energy decreases from $E_0$
down to $E$. This means that the transverse normalized emittances have
decreased: $ \EN = \gamma \E = \EN_0(E/E_0)$. One can reaccelerate the
electron beam up to the initial energy and repeat the procedure. Then
after N stages of cooling $ \EN /\EN _0 = (E/E_0)^N$ (if \EN\ is far
from its limit).

The ultimate emittance can be estimated in the following way. In the
electron beam reference system the counter moving laser photons have
an energy $\omega \sim \gamma\omega_0$ ($\gamma=E/mc^2$ and $\omega_0$
is the energy of laser photons) and scatter almost isotropically
(roughly) without change of the energy. As a result, after multiple
Compton scattering, the transverse energy of the electrons is equal to
the transverse energy of laser photons
\begin{equation}
p_t^2/m \sim \gamma \omega_0.
\label{in1}
\end{equation}
Here we assumed that $\gamma\omega_0 \ll mc^2$. On the other hand the
r.m.s.  angular spread of the electrons in the laboratory system by
definition (see above)
\begin{equation}
\frac{p_t}{\gamma mc} \equiv \sqrt{\frac{\EN}{\gamma \beta}},
\label{in2}
\end{equation}
where \EN\ is the normalized emittance and $\beta$ is the
beta-function of the electron beam in the cooling region. Substituting
eq.\ref{in1} to eq.\ref{in2} we obtain
\begin{equation}
\EN\ \sim \frac{\lambda_C}{\lambda}\beta,
\label{in3}
\end{equation}
where $\lambda_C= \hbar/mc$ is the Compton wave length, $\lambda$ is
the laser wave length.

The physics of the cooling process is almost the same as radiative
cooling of electrons in damping rings. However, here the process takes
only 1 ps and the ultimate emittance is much lower than that in the
damping rings. This is because in the ``linear'' laser cooling there
are no bends (as in damping rings) which cause a growth of the
horizontal emittance. Also the intra-beam scattering is not important
due to a short ``damping'' time and following fast acceleration.

There are several question to this method 
\begin{enumerate}
\item requirements for laser parameters (these parameters should be
attainable);

\item an energy spread of the beam after cooling (at the final energy
  of a linear collider it is necessary to have $\sigma_E/E \sim 0.1
  \%$; also with a large energy spread it is difficult to ``match''
  the cooled beam  with the accelerator due to
  the chromaticity of focusing systems and also to focus the beams for the
  next stage of cooling.)

\item the limit on the final normalized emittances (it is desirable to
  have this limit lower than that obtained with storage rings and
  photoguns);

\item depolarization of electron beams (polarization is very important for
linear colliders).

\item radiation damage of mirrors by the photons scattered at the large angles
  (these are X-ray photons).

\end{enumerate}

Below we will see that this method is perfectly suited for linear
colliders, however the main problem here obtaining of a very powerful
laser pulses with high repetition rate. On my opinion, the most
promising approach to this problem is a pulse stacking of laser pulses
in an optical cavity~\cite{Tfrei},\cite{Tsit2}. All these problems and
possible solutions are discussed below.

\section{Flash energy.}
In the cooling region, a laser photon with  energy $\omega_0$ (wave
length $\lambda$) collides almost head--on with an electron of
energy $E$. The kinematics is determined by two parameters $x$ and
$\xi$ \cite{GKST83,TEL90,TEL95}. The first one
\begin{equation}
x=\frac{4E \omega_0}{ m^2c^4}= 0.019\left[\frac{E}{\GEV}\right]
\left[\frac{\MKM}{\lambda}\right]
\label{x}
\end{equation}
determines the maximum energy of the scattered photons:
\begin{equation}
\omega_m=Ex/(x+1) \sim 4\gamma^2\omega_0 \;\;\;\;(x \ll 1).
\label{om}
\end{equation}
If the electron beam is cooled at the initial energy $E_0 = 5\; \GEV$
(after damping ring and bunch compression) and $\lambda = 0.5\; \MKM$
(Nd:glass laser) then $x_0 \simeq 0.2$ (we will provide $E$ and $x$
with the index 0 for designation of their values at the begining of a
cooling region).

For our further consideration we will need also the following formulae
for the Compton scattering in the case $x<<1$.  The energy spectrum of
scattered photons(normalized per one scattering)~\cite{GKST83} 
\begin{equation}
dp = (3/2)[1-2\omega/\omega_m +2(\omega/\omega_m)^2]d\omega/\omega_m.
\label{spectrum}
\end{equation}
 The angle of the electron after scattering  
\begin{equation}
\theta_1^2 = (\omega_m\omega - \omega^2)/(\gamma^2E^2).
\label{angle}
\end{equation}
The second parameter characterizes a strength of an electromagnatic
wave \footnote{Usually $\xi^2$ is defined with $\bar{B^2}$ instead of
  $B_0^2$, that is 2 time smaller than in my ``definition'' given in
  ref.\cite{TSB1}, which I use here also.}
\begin{equation}
\xi^2 = \left(\frac{eB_0\hbar}{m\omega_0 c}\right)^2,
\label{xi2}
\end{equation}
where $B_0$ is the  magnetic (or electric) field strength in the
laser wave.  At $\xi^{2} \ll 1$ an electron interacts with one
photon from the field (Compton scattering, undulator radiation), while
at $\xi^2 \gg 1$ an electron scatters on many laser photons
simultaneously (synchrotron radiation (SR), wiggler). We will see that
in the considered method $\xi^2$ may be ``small'' and ``large''.

In the cooling region near the laser focus the r.m.s radius of the
laser beam depends on the distance $z$ to the focus (along the beam)
in the following way~\cite{GKST83}: $ r_\gamma = a_\gamma \sqrt{ 1 +
z^2 /Z_R^2}$, where $Z_R =2\pi a^2_\gamma /\lambda$ is the Rayleigh
length (an effective depth of laser focus), $a_\gamma$ is the
r.m.s. focal spot radius.  The density of laser photons is $n_\gamma =
(A/\pi r^2_\gamma \omega_0) \exp (-r^2/r^2_\gamma) F_\gamma(z+ct)$,
where $A$ is the laser flash energy and $\int F_\gamma(z) dz = 1.$

In the case of strong field ($\xi^2 \gg 1$) it is more appropriate to
speak in terms of strength of the electromagnetic field which is
$\bar{B^2}/4\pi=n_{\gamma}\omega_0, B=B_0cos(\omega_0t/\hbar-kz)$.
Assuming $F_{\gamma}=1/l_{\gamma}$ and $Z_R \ll l_{\gamma} \simeq l_e$
and using the classical formula for radiation loss
($dE/dx=(2/3)r^2_e\gamma^2 B^2,\;r_e=e^2/mc^2$) we obtain the ratio of
emittances before and after the laser target
\begin{equation}
 \frac{\EN_0}{\EN} \simeq \frac{E_0}{E} = 1+\frac{r_e^2}{3m^2c^4} \int
B_0^2dz = 1+\frac{64\pi^2r_e^2\gamma_0}{3mc^2\lambda\l_e}A \
\label{en0}
\end{equation}
\begin{equation}
 A[J] = \frac{25\lambda[\MKM\ ]l_e[\MM\ ]}{E_0[\GEV\ ]}
\left(\frac{E_0}{E}-1\right).
\label{A}
\end{equation}
These equations are correct at $x \ll 1$ for any value
of $\xi^2$.  For example: at $\lambda=0.5\; \MKM,\; l_e=0.2\; \MM,
E_0=5\;\GEV, E_0/E=10$ the required laser flash energy $A = 4.5\; J.$
To reduce the laser flash energy in the case of long electron bunches,
one can compress the bunch (length) before cooling as much as possible
and stretch it after cooling up to the required value.

The eqs.\ref{en0},\ref{A} were obtained for $Z_R \ll l_{\gamma} \sim
l_e$ and give the minimum flash energy for a certain ratio $E_0/E$.
To further estimate  the photon density at the laser focus we
will assume $Z_R \sim 0.25l_e$. In this case, the required
flash energy is still close to its minimum, but the field strength
is not so high as for very small $Z_R$. From the previous
equation for $Z_R = 0.25 l_e$ it follows $B_0^2/(8\pi) =
\omega_0n_{\gamma}=A/(\pi a_{\gamma}^2 l_e) =8A/(\lambda l_e^2)$.
Substituting $B_0$ into eq.\ref{xi2} we get
\begin{equation}
\xi^2 = \frac{16r_e\lambda A}{\pi l_e^2 mc^2} =
\frac{3\lambda^2}{4\pi^3 r_e l_e \gamma_0}\left(\frac{E_0}{E}-1\right
) \ =  = 4.3 \frac{\lambda^2 [\MKM\ ]}{l_e [\MM\ ]E_0 [\GEV\ ]}
\left(\frac{E_0}{E}-1\right ).
\label{xi22}
\end{equation}  
Example: for $\lambda$ = 0.5 \MKM, $E_0$ = 5 \GEV\ , $ E_0/E$ = 10,
$l_e$ = 0.2 \MM\ (the NLC project) $\Rightarrow \xi^2$ = 9.7. For
larger bunch lengths and shorter wave lengths, $\xi^2$ may be smaller.
So, both ''undulator'' and ''wiggler'' cases are possible.

Later we will see that in order to have lower limit on emittance and
smaller depolarization it is necessary to have a low $\xi^2$. With a
usual optics one can reduce $\xi^2$ only by increasing $l_{\gamma}$
(and $Z_R$) with a simultaneous increase of the laser flash
energy. From (\ref{A}) and (\ref{xi22}) we get
\begin{equation}
A \propto \frac{\lambda^3}{\gamma_0^2 \xi^2}\left(\frac{E_0}{E}-1 \right)^2.
\end{equation}
   
Is it possible to reduce $\xi^2$ keeping all other parameters
(including flash energy) constant?  Yes, providing a way to stretch
the focus depth without changing the radius of this area is found. In
this case, the collision probability (or $\int B^2dz$) remains the same
but the maximum value of $\xi^2$ will be smaller.  A solution of this
problem was given in \cite{TSB1}. It is based on the
non-monochromaticity of the laser light and the chirped pulse
technique. In this scheme, the cooling region consists of many laser
focal points (continuously) and light comes to each point exactly at
the moment when the electron bunch is there.  One can consider that a
short electron bunch collides on its way sequentially with \n\ 
(``number of focuses'') short light pulses of  length $l_{\gamma}
\sim l_e$ and focused with $2Z_R \sim l_e$.  There is one 
restriction on \n: along the cooling length $L \approx \n\cdot l_e$ the
transverse size of an electron beam should be smaller than the laser
spot size $a_{\gamma} \simeq \sqrt{\lambda Z_R/2\pi} \sim
\sqrt{\lambda l_e/4\pi}$.  In further examples we will use $\n\ \sim 10$
for stretching the cooling region from 100 \MKM\ to 1 \MM.

Other method of ``stretching'' is using of several lasers focused to
different, though this require larger flash energy (see sect.\ref{system}).

\section{Energy spread}
The electron energy spread   arises from the 
quantum-statistical nature of radiation.  After energy loss 
$\Delta E$, the increase of the energy spread
$\Delta(\sigma_E^2)=\int\varepsilon^2 \dot{n}(\omega) d\omega dt
=-aE^2\Delta E$,  where $\dot{n}(\omega)$ is the spectral density
of photons emitted per unit time, $a=14\omega_0/5m^2c^4=7x_0/10E_0$ for
the Compton case and $a=55\hbar e B_0/(8\pi\sqrt{3}m^3c^5)\;$ for the
``wiggler'' case.
   
There is the  second effect which leads to  decreasing  the energy
spread.  It is due to the fact that $dE/dx \propto E^2$ and an electron
with higher (lower) energy than the average  loses more (less) than on
average.  This results in the damping: $d(\sigma_E^2)/\sigma_E^2 =
4dE/E $ (here $dE$ has negative sign). The full
equation for the energy spread is $d\sigma_E^2 = -aE^2dE +
4(dE/E)\sigma_E^2$, with  solution
$\sigma_E^2/E^2 = \sigma_{E_0}^2E^2/E_0^4+
aE_0(E/E_0)(1-E/E_0)$. In our case
\begin{equation}
\frac{\sigma_E^2}{E^2}  \sim\frac{\sigma_{E_0}^2E^2}{E_0^4}+
  \frac{7}{10}x_0(1+\frac{275\sqrt{3}}{336\pi}\xi)\frac{E}{E_0}
  \left(1-\frac{E}{E_0}\right),
\label{c7}
\end{equation}
here the results for the Compton scattering and SR are joined together.
Example: at $\lambda=0.5\;\MKM,\; E_0=5\; \GEV\ (x_0 = 0.19)$ and
$E_0/E=10$,  the  Compton term alone gives $\sigma_E/E \sim 0.11$
and with the ``wiggler'' term ($\xi^2 = 9.7$, see the example above)
$\sigma_E/E \sim 0.17$.
What $\sigma_E/E$ is acceptable? In the last example $\sigma_E/E \sim
0.17$ at E = 0.5 GeV (after  cooling). This means that at the
collider energy E = 250 GeV we will have $\sigma_E/E \sim 0.034\%$,
that is better than necessary (about 0.1 \%).

In a two stage cooling system, after reacceleration to the initial
energy $E_0$ = 5 GeV the energy spread is $\sigma_E/E_0 \sim 1.7\%$.
For this value there may be a problem with focusing of electrons which
can be solved using a focusing scheme with correction of chromatic
aberrations. It is even more difficult to ``match'' the electron beam
after the laser cooling with the accelerator (see
sect.\ref{radiation}.
\section{Minimum normalize emittance.}

It is determined by the quantum nature
of the radiation. Let us start with the case of  pure Compton
scattering at $\xi^2 \ll 1$ and $x_0 \ll 1$. In this case, the
scattered photons have the energy distribution given by eq.\ref{spectrum}
The angle of the electron after scattering is connected with the energy by 
eq.\ref{angle}. After averaging over the
energy spectrum we get the average $\theta_1^2$ in one collision:
$\langle\theta_1^2\rangle = 12\omega_0^2/(5m^2c^4)$. After many Compton
collisions ($N_{coll}$) the r.m.s. angular spread in i=x,y projection
$\Delta\langle\theta^2_i\rangle = 0.5\Delta\langle\theta^2\rangle =
0.5N_{coll}\langle\theta_1^2\rangle = -0.5(\Delta
E/\bar{\omega})\langle\theta_1^2\rangle = -3\omega_0\Delta E/5E^2$.

The normalized emittance $\ENI{^2} = (E^2/m^2c^4)\langle r_i^2\rangle
\langle\theta_i^2\rangle$ does not change when $\Delta
\langle\theta_i^2\rangle/ \langle\theta_i^2\rangle$ = $-2\Delta E/E.$
Taking into account that
$\langle\theta_i^2\rangle\equiv\ENI\ /\gamma\beta_i$ we get the
equilibrium emittance due to the Compton scattering
\begin{equation}
\ENI_{, min} \approx
 0.5\gamma E \beta_i\Delta \langle\theta_i^2\rangle/\Delta E =
 \frac{3\pi}{5}\frac{\lambda_c}{\lambda}\beta_i 
 =\frac{7.2\times 10^{-10}\beta_i [mm]}{\lambda [\MKM\ ]} \;\;
  \mbox{m$\cdot$rad},
\label{ecompt}
\end{equation}
where $\lambda_c=\hbar/mc$.  For example: $\lambda =0.5\; \MKM,
\;\beta=l_e/2=0.1\;\MM\ ($NLC$) \Rightarrow
\EN_{,min}=1.4\times10^{-10}\;\M\cdot$rad. For comparison in the NLC
project the damping rings have $\ENX\ =3\times 10^{-6}\; \M\cdot$rad,
$\ENY\ =3\times 10^{-8}\; \M\cdot$rad.

If $\xi^2 \gg 1$,  the electron moves as
in a  wiggler.Assume that the ``laser wiggler'' is planar and deflects the
electron in the horizontal plane. If an electron with  energy E
emits a photon with energy $\omega$ along its trajectory the
emittance changes as follows \cite{WID}: $\delta \EX =
(\omega^2/2E^2)H(s);\; H(s)=\beta_x\eta_x^{\prime 2}
+2\alpha_x\eta_x\eta_x^{\prime} + \gamma_x\eta_x^2$; where $\alpha_x =
-\beta_x^{\prime}/2,\; \gamma_x = (1+\alpha_x^2)/\beta_x,\; \beta_x $
is the horizontal beta-function, $\eta_x$ is the dispersion function,
$s$ is the coordinate along the trajectory. For $\beta_x = const$ the
second term in {\it H} is equal to zero, the second term in a wiggler
with $\lambda_w \ll \beta$ is small, so that {\it
H(s)} $\approx \beta\eta^{\prime 2}$. In a sinusoidal wiggler field
$B(z)=B_wcos\;k_wz,\; k_w=2\pi/\lambda_w$,
$\eta^{\prime\prime}=1/\rho$ ($\rho$ is the radius of curvature) one
finds that $\eta^{\prime} = (eB_w/k_wE)\; sin\;k_wz$.  The increase
of \EX\ on a distance $dz$ is
\begin{equation}
\Delta \E_x = \int \frac{H}{2}\left(\frac{\omega}{E}\right)^2
\dot{n}(\omega)d\omega dt = \frac{55}{48\sqrt{3}}\frac{r_e\hbar
c}{(mc^2)^6} E^5 \langle \frac{{\it H}}{\rho^3}\rangle dz,
\label{De}
\end{equation}
where $ \langle {\it H}/\rho^3\rangle_w =
8\beta_x\lambda_w^2(eB_w)^5/(140E^5\pi^3)$ for the wiggler and
$\dot{n}(\omega)$ is the spectral density of photons emitted per unit
time.  The energy loss averaged over the wiggler period is $\Delta E =
r_e^2B_w^2E^2dz/(3m^2c^4)$. The normalized emittance $\EN=\gamma\E\ $
is not changed when $Ed\E\ + \E\ dE =0$. Using this
and replacing $B_w$ by 2$B_0$, $\lambda_w$ by $\lambda/2$ we obtain
the equilibrium normalized emittance in the linear polarized
electromagnatic wave for $\xi^2 \gg 1$
\begin{equation}
\ENX = \frac{11e^3\hbar c \lambda^2 B_0^3
\beta_x}{24\sqrt{3}\pi^3(mc^2)^4} =
\frac{11}{3\sqrt{3}}\frac{\lambda_C}{\lambda}\beta_x \xi^3 \approx
\frac{8\cdot 10^{-10}\beta_x [\MM\ ] \xi^3}{\lambda [\MKM\ ]} \;\;
\mbox{m$\cdot$rad}.
\label{ex}
\end{equation}
Using eq.\ref{xi22} we can get a scaling of the minimum \ENX\ for a multistage
cooling system with a cooling factor $E_0/E$ in one stage: $\ENX\ 
\propto \beta_x\lambda^2(E_0/E)^{3/2}/(l_e\gamma_0)^{3/2}$ when
$l_{\gamma} \sim l_e$(minimum A) and $\ENX\ \propto
\beta_x\lambda^{7/2}(E_0/E)^{3}/(\gamma_0^3A^{3/2})$ for free A and
$\l_{\gamma} > l_e$ (for $\beta_x = const$).  Stretching the laser
focus depth by a factor \n\ , one can further reduce the horizontal
normalized emittance: $\ENX\ \propto 1/\n^{1/2}$(if $\beta_x \propto
\n$).  For our previous example we have $\xi^2 =9.7$ and $\ENX\ =
5\cdot 10^{-9}\;\M\cdot$rad (in the NLC \ENX\ = 3$\cdot 10^{-6}$
m$\cdot$rad). Stretching  the cooling region with \n=10, further
decreases the horizontal emittance by a factor 3.2.

Comparing with the Compton case (\ref{ecompt}) we see that in the strong
field the horizontal emittance is larger by a factor
$\xi^3$. The origin of this factor is clear: $\ENX \propto
\eta_x^{\prime 2} \omega_{crit.}$, where $\eta_x^{\prime} \sim
\xi\theta_{compt.}$ and $\omega_{crit.} \sim \xi\omega_{compt.}$.

Let us  roughly estimate the minimum vertical normalized emittance at
$\xi \gg\ 1$. Assuming that all photons are emitted at an angle
$\theta_y = 1/(\sqrt{2}\gamma)$ with the $\omega = \omega_c$ similarly
to the Compton case, one  gets $\Delta\langle\theta^2_y\rangle =
(\omega_c \Delta E)/(2\gamma^2E^2) = -(3e\hbar \bar{B_w}\Delta
E)/(4E^2mc)$.  Using the first part of eq.\ref{ecompt} we get 
\begin{equation}
\ENY_{min} \sim \frac{3}{8}\frac{\hbar e \bar{B_w}}{m^2c^3}\beta_y =
 3 \left(\frac{\lambda_C}{\lambda}\right)\beta_y\xi \approx
 \frac{1.2\cdot 10^{-9}\beta_y [\MM\ ] \xi}{\lambda [\MKM\ ]} \;\;
\mbox{m$\cdot$rad}.
\label{ey}
\end{equation}   
For the previous example (NLC beams), eq.\ref{ey} gives $\ENY_{min} \sim
7.5\cdot 10^{-10}$ m$\cdot$rad (for  comparison in the NLC project
$\ENY=3\cdot10^{-8}$ m$\cdot$rad). The scaling: $\ENY\ \propto
\beta_y(E_0/E)^{1/2}/(l_e\gamma_0)^{1/2}$ when $l_{\gamma} \sim
l_e$(minimum A) and $\ENY\ \propto
\beta_y\lambda^{1/2}(E_0/E)/(\gamma_0A^{1/2})$ for free A and
$\l_{\gamma}>l_e.$

For arbitrary $\xi$ the minimum emittances can be estimated as the sum
of (\ref{ecompt}) and (\ref{ex}) for \ENX\ and sum of (\ref{ecompt})
and (\ref{ey}) for \ENY\ 

\begin{equation}
  \ENX \approx
  \frac{3\pi}{5}\frac{\lambda_C}{\lambda}\beta_x(1+1.1\xi^3);\;\; \ENY
  \sim \frac{3\pi}{5}\frac{\lambda_C}{\lambda}\beta_y(1+1.6\xi).
\label{efin}
\end{equation}

\section{Depolarization}
Finally, let us consider the problem of the depolarization. For the
Compton scattering the probability of spin flip in one collision is
$w=(3/40)x^2$ for $x \ll 1$ (it follows from formulae of
ref.\cite{KOT}). The average energy losses in one collision are
$\bar{\omega} = 0.5xE$. The decrease  of polarization degree
 after many collisions is $dp = 2wdE/\bar{\omega} = (3/10)x(dE/E) =
(3/10)x_0(dE/E_0)$.  After integration, we obtain the relative decrease
of the longitudinal polarization $\zeta$ during one stage of the  cooling
(at $E_0/E \gg 1$)
\begin{equation}
        \Delta\zeta/\zeta = 0.3x_0\;\;\; \propto E_0/\lambda,
\label{dz1}
\end{equation}
For $\lambda = 0.5\; \MKM\ $ and $E_0 = 5\; \GEV\ $, we have $x_0 =
0.19$ and $\Delta\zeta/\zeta = 5.7 \%$. This is valid only for $\xi^2
\ll 1.$

In the case of strong field ($\xi^2 \gg 1$) the spin flip probability
per unit time is the same as in the uniform magnetic field \cite{LAN}
$w = (35\sqrt{3}r_e^3\gamma^2ce\bar{B^3})/(144\alpha(mc^2)^2)$,
where for the wiggler $\bar{B^3} = (4/3\pi)B_w^3$. Using the relation
between $dE$ and $dt$ in the wiggler we get
\begin{equation}
  \frac{\Delta\zeta}{\zeta} =
  \int\frac{35\sqrt{3}er_eB_0}{9\pi\alpha(mc^2)^2}dE \sim
  \frac{35\sqrt{3}}{36\pi}x_0\xi.
\label{dz2}
\end{equation}
For the general case, the depolarization can be estimated as the sum of
equations (\ref{dz1}) and (\ref{dz2})
\begin{equation} 
 \Delta\zeta/\zeta \approx 0.3x_0(1+1.8\xi). 
\label{dzfin}
\end{equation}
For the previous example with $\xi^2 =9.7$ and $x_0 = 0.19$ we get
$\Delta\zeta/\zeta = 0.057+0.32 = 0.38$, that is not acceptable. This
example shows that the depolarization effect imposes very
demanding requirements on the parameters of the cooling system. The
main contribution to depolarization gives the second term. 
Stretching the cooling region by a factor of ten
we  can get $\Delta\zeta/\zeta = 0.057+0.1\sim 15\%$.

\section{Ponderomotive forces.}
  
It is well known that in a non-uniform oscillating field an avarage
force acting on a particle is non-zero, this is so called {\it
  ponderomotive} force.  This force leads to the repulsion of the
electrons from the laser focus. This effect was not described in my
first paper on laser cooling~\cite{TSB1}, but it was checked that it
is not essential for the considered examples. Nevertheless, it is
important, especially for low beam energies, let us consider this
effect in more details.

The total force acting on an electron colliding head-on with a laser
wave~\footnote{In this section $\omega_0$ is the frequency, in other
  sections it is the energy of the laser photon.}
\begin{equation}
F=2eE_0(x)\sin{2\omega_0 t}=2e(E_0+\frac{\partial E_0}{\partial x}x)
 \sin {2\omega_0 t}. 
\label{pond}
\end{equation}
Substituting $x \approx -(eE_0/2\gamma m \omega_0^2)\sin{2\omega_0 t}$ we get
\begin{equation}
F=2e(E_0-\frac{e}{2\gamma m \omega_0^2}\frac{\partial E_0}{\partial x}
E_0\sin{2\omega_0 t}) \sin {2\omega_0 t}. 
\label{pond1}
\end{equation}
After averaging over time we get the ponderomotive force
\begin{equation}
\bar{F}=-\frac{e^2}{4\gamma m \omega_0^2}\frac{\partial E_0^2}{\partial x}=
-\frac{\partial U_{eff}}{\partial x}, 
\label{pond2}
\end{equation}
where the effective potential 
\begin{equation}
U_{eff}=\frac{mc^2}{4\gamma}\xi^2.
\label{pond3}
\end{equation}
Particle move in the direction with the minimum ponential, in our case
the electrons are repulsing from the laser target. Note, that in all
considerations we assume a linearly polarized laser light with $\vec{E}$
laying in the horizontal plane. In this case there are forces only in the
horizontal direction.

Let us assume that the laser spot size is about a factor of 2 larger
than the horizontal size of the electron beam in the laser focus, then $
\partial \xi^2/\partial x \sim \xi^2/2\sigma_{x,L}$. The deflection angle
on the cooling length $l_c$
\begin{equation}
\Delta \vartheta \sim \frac{p_t}{p} \sim \frac{F(l_c/c)}{\gamma m c}\sim
\frac{\xi^2l_c}{8\gamma^2\sigma_{x,L}}.
\label{pond4}
\end{equation}
The ponderomotive forces are not important if this angle is smaller
than the angular spread of the electrons in the cooling region:
$\Delta \vartheta < \sqrt{\ENX\ /\gamma\beta}$. This gives the minimum
normalized horizontal emittance when ponderomotive forces are still
not important
\begin{equation}
\ENX_{,min} \sim \frac{\xi^4 l_c^2}{64\gamma^3\sigma^2_{x,L}}\beta.
\label{pond5}
\end{equation}
We have seen before that the cooling by a factor of ten can be done at
$\xi^2 \sim 1$ on the length $l_c \sim$ 1 mm. The minimum laser spot
size $\sigma_{x,L} \sim \sqrt{\lambda Z_R/4\pi} \sim 2.5$ \MKM\ at $\lambda
= 0.5$ \MKM\ and $Z_R \sim \sigma_{z,e} \sim 0.1$ mm. Note, that for
the first stages of cooling after the damping ring the minimum
$\sigma_{x,L}$ is determined by the electron beam size and should be
larger by a factor of 3 than this estimate. Now we investigate the
minimum emittance therefore let us take $\sigma_{x,L}=2.5$ \MKM. Substituting
this number to eq.\ref{pond5} and we get for  $E=500$ \GEV\ (the minimum
energy in the cooling process)  the estimate of the minimum
normalized emittance when ponderomotive forces are still not important
\begin{equation}
\ENX_{min} \sim 2.5\times 10^{-9} \beta[\MM]\;\; \mbox{m$\cdot$rad}
\label{pond6}
\end{equation}
This is exectly equal to the limit on the emittance in the laser
cooling obtained in the previous sections.  So, at the chosen beam energies the
ponderomotive forces still do not limit the minimum emittance in the
laser cooling, but they will be important at the lower beam energies.

\section{Some "intermediate" conclusions.} 

Before considering ``technical'' aspects in the laser cooling we can
summarize the results of the previous section as a possible
"optimistic" set of parameters for the laser cooling:
$E_0 = 4.5$ GeV, $l_e=0.2 $ mm, $\lambda = 0.5$ \MKM, flash energy $A
\sim 10 $ J, focusing system with stretching factor \n=10. The final
electron bunch will have an energy of 0.45 \GEV\ with an energy spread
$\sigma_E/E \sim 13 \%$, the normalized emittances \ENX,\ENY\ are
reduced by a factor 10, the limit on the final emittance is $\ENX\
\sim\ENY\ \sim 2\times 10^{-9}\;$ m$\cdot$rad at $\beta_i= 1\; \MM$,
depolarization $\Delta\zeta/\zeta \sim 15\%$.  
  The two stage system with the same parameters gives
100 times reduction of emittances (with the same limits).

For the cooling of the electron bunch train one laser pulse can be
used many times.  According to (\ref{A}) $\Delta E/E = \Delta A/A$ and even
25\% attenuation of laser power leads only to small additional energy
spread.

\section{Laser systems}

We have seen that the ``very'' minimum flash energy required for the
one stage of the laser cooling is about 10 J for visible light and
$E\sim 5$ GeV beam energy. If $\lambda \sim 1$ \MKM\ (most powerful
solid state lasers) then $A\sim 20$ J. If the electron beam is
somewhat longer than in the considered examples, say $\sigma_z=$
300-400 \MKM\ as in the TESLA project, then the required flash is
already $\sim$ 40--50 J. Beside, as we will see in the next section,
for decreasing the radiation damage of the mirrors one has to put the
mirrors far enough from the electron beam.  That leads already to more
than 100 J flash energies. Moreover, the repetition rate should be
equal to the rep.rate of linear colliders which is of the order of 10
kHz.  So, the average power of the laser system should be of the order
of one MegaWatt! At present, the best short pulse laser systems can
produce several Joule pulses with the repetition rate several Hz and
only there are hopes that next year a commercial 100 W laser with
short (ps) pulses will be built~\cite{NLC},\cite{Perry}. However, the
situation is not so pessimistic.

To overcome the ``repetition rate'' problem it is quite natural to
consider a laser system where one laser bunch is used for e$\to
\gamma$ conversion many times. Indeed, one Joule laser flash contains
about $10^{19}$ laser photons and only $10^{12}$ photons are knocked
out in the collision with one electron bunch ($\sim 100$ Compron
scattering per one electron).

The simplest solution is to trap the laser pulse to some optical loop and
use it many times.~\cite{NLC} In such a system the laser pulse enters
via the film polarizer and then is trapped using Pockels cells and
polarization rotating plates.  Unfortunately, such a system will not
work with Terawatt laser pulses due to a self-focusing effect.

Fortunately, there is one way to ``create'' a powerful laser pulse in
the optical ``trap'' without any ``nonlinear'' material inside (only very thin
dielectric coating of  mirrors). This very promising technique is
discussed below.

\subsection{Laser pulse stacking in an ``external'' optical cavity.} 

Shortly, the method is the following. Using the train of low energy
laser pulses one can create in the external passive cavity
(with one mirror having some small transparency) an optical pulse of
the same duration but with much higher energy (pulse stacking). This
pulse circulates many times in the cavity each time colliding with
electron bunches passing the center of the cavity.

The idea of pulse stacking is simple but not trivial and not well
known. This method is used now in several experiments on detection of
gravitation waves.  In my opinion, pulse stacking is very natural for
photon colliders and allows not only to build a relatively cheap laser
system for $e\to\gamma$ conversion but gives us the practical way for
realization of laser cooling, i.e. opens up the way to ultimate
luminosities of photon colliders.

\begin{figure}[!htb]
\centering
\vspace*{0.7cm} 
\hspace*{-0.2cm} \epsfig{file=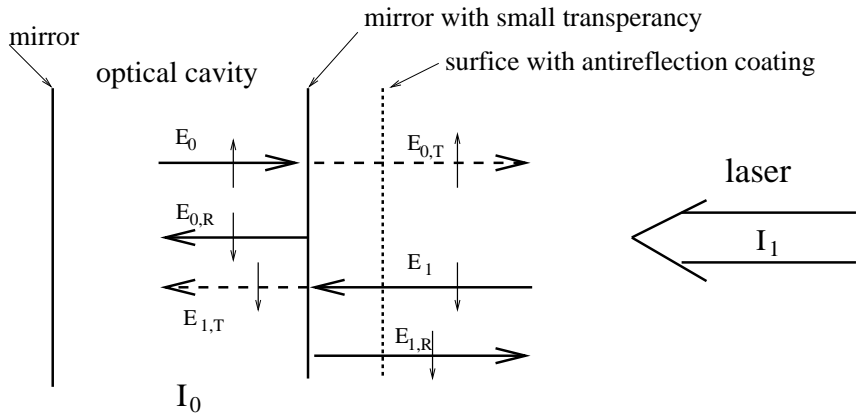,width=11.5cm,angle=0} 
\vspace*{-0.cm} 
\caption{Principle of pulse stacking in an external optical cavity.}
\vspace{2mm}
\label{cavity}
\vspace{5mm}
\end{figure}

As this method is very important (may be crucial) for the laser
cooling, let me explain it in more detail~\cite{Tfrei},\cite{Las1}.
The principle of pulse stacking is shown in Fig.\ref{cavity}.
The secret consists in the following. There is a well known optical
theorem: at any surface, the reflection coefficients for light coming
from one and the other sides have opposite signs. In our case, this means
that light from the laser entering through semi-transparent mirror into
the cavity interferes with reflected light inside the cavity {\bf
constructively}, while the light leaking from the cavity interferes
with the reflected laser light {\bf destructively}. Namely, this fact produces
asymmetry between cavity and space outside the cavity!  

Let R be the reflection coefficient, T the transparency coefficient
and $\delta$ the passive losses in the right mirror. From the energy
conservation $R+T+\delta =1$. Let $E_1$ and $E_0$ be the amplitudes
of the laser field and the field inside the cavity. In equilibrium,
$E_0= E_{0,R} + E_{1,T}$. 
Taking into account that $E_{0,R}=E_0\sqrt{R}$, $E_{1,T}=E_1\sqrt{T}$ and 
$\sqrt{R}\sim 1-T/2-\delta/2$ for $R\approx 1$ we obtain
$E_0^2/E_1^2=4T/(T+\delta)^2.$ 
The maximum ratio of intensities is obtained at $T=\delta$, then 
$I_0/I_1=1/\delta \approx Q$,
where $Q$ is the quality factor of the optical cavity.  Even with two
metal mirrors inside the cavity, one can hope to get a gain factor of about
50--100; with multi-layer mirrors it can reach $10^5$. ILC(TESLA)
colliders have 120(2800) electron bunches in the train, so the factor
100(1000) would be perfect for our goal, but even the factor of ten
means a drastic reduction of the cost.

   Obtaining of high gains requires a very good stabilization of cavity
size: $\delta L \sim \lambda/4\pi Q$, laser wave length: $\delta
\lambda/\lambda \sim \lambda/4\pi QL$ and distance between the laser
and the cavity: $\delta s \sim\lambda/4\pi$. Otherwise, the  condition of
constructive interference will  not be fulfilled. Besides, the
frequency spectrum of the laser should coincide with the cavity modes,
that is automatically fulfilled when the ratio of the cavity length and
that of the laser oscillator is equal to an integer number 1, 2, 3... . 

For $\lambda = 1\;\mu m$ and $Q=100$, the stability of the cavity
length should be about $10^{-7}$ cm. In the LIGO experiment on
detection of gravitational waves which uses similar techniques with
$L\sim 4$ km and $Q\sim 10^5$ the expected sensitivity is about
$10^{-16}$ cm.  In comparison with this project our goal seems to be
very realistic.

      In HEP literature I have found only one reference on pulse
stacking of short pulses ($\sim 1$ ps) generated by FEL~\cite{HAAR}
with the wave length of 5 $\mu$m. They observed pulses in the cavity
with 70 times the energy of the incident FEL pulses, though no long
term stabilization was done.

\section{Radiation damage of mirrors and other ``technical'' 
aspects \label{radiation}}

The use of pulse stacking in the optical cavity makes the idea of
laser cooling quit realistic.

Considering a practical scheme for laser cooling we should take into
account many important practical aspects: 

$\bullet$ Radiation damage of the mirrors. X-ray radiation due to the
Compton scattering here is  many orders larger than the radiation level
at the same angles in the $e \to \gamma$ conversion point.  It is so
because a) the electron energies are lower and b) each electron undergoes
about one hundred Compton scattering.  At $\vartheta \gg 1/\gamma$ and $x
\ll 1$ ($x$ is defined in sect.2) the energy of the Compton scattered
photons $\omega = 4\omega_0/\vartheta^2$ and does not depend on the
electron energy.~\cite{GKST83} However, at the lower beam energies the
spectrum is softer ($\omega_{max} = 4\omega_0\gamma^2)$ and more
photons (per one Compton scattering) have large angles. Simple
calculations show that the number of photons/per 
electron emitted on the angle $\vartheta$ during the cooling of electrons
from some large energy to the energy $E_{min}$ is 
$$
dn/d\Omega =mc^2/4\pi\omega_0\gamma_{min}^3\vartheta^4.
$$

The total energy  hitting the mirrors/cm$^2$/sec is 
$$
dP/dS=mc^2N\nu/\pi\gamma_{min}^3\vartheta^6 L^2,
$$
where $L$ is the distance between the collision (cooling) point
(CP) and the focusing mirrors, N and $\nu$ are the number of electrons in
the bunch and the collision rate. One can see a strong dependence of X-ray
background on $\gamma_{min}$ and $\vartheta$. During the cooling the
electron beam loses almost all its energy to photons. For $E_0=5$ GeV,
$N=2\times 10^{10}$, $\nu = 15$ kHz the total energy losses are about
200 kW, fortunately the flux decreases rapidly with increasing the angle. At
$\vartheta$ = 30 mrad and $L=5$ m the power density $dP/dS \sim
10^{-5}$ W/cm$^2$ and X-ray photons have an energies of about 4 keV (for 1
\MKM\ laser wave length). My estimations shows that rescattering of photons on
the quads can give a comparable background. 

I have describing this item in detail because for laser cooling the
required flash energy is very high and to reach the goal we need very
high reflectivity of the mirrors in the optical cavity. For TESLA with
3000 bunches in a train it would be nice to have mirrors with
$R>0.999$. Such values of R are not a problem for dielectric mirrors,
however the radiation damage may cause problems, better to avoid this
problem.

$\bullet$ Laser spot size should be several times larger than that of
the focused electron beam to avoid an additional energy spread of
the cooled electrons.

$\bullet$ The cooled electron beam at the energy E=500--1000 GeV has an
energy spread of $\sigma_E/E \sim 15$ \% at the point where the
$\beta$- function is small ($\sim 1-5$ mm). Matching
this beam with the accelerator is not a simple problem and requires
special insertions for chromaticity correction. A similar
problem exists for the final focus at linear colliders, it has been
solved and tested at the FFTB at SLAC. Here the factor
$(F/\beta)\sigma_E/E$ characterizing the chromaticity problem is
smaller and the beam energy is 500 times smaller, so one can hope that
it will be no problem.

$\bullet$ The parameter $\xi^2$ (defined above) should be small enough
($\le$ 1) to keep the minimum attainable emittance, depolarization and
the energy spread small enough. This is impossible with one laser
(with required flash energy) without additional "stretching" of the
cooling region along the beam line.  The simplest way to do this is to
focus several lasers at different points along the beam axis.

\section{Possible variant of the laser cooling system \label{system}}

\begin{figure}[!htb]
\centering
\vspace*{-0.0cm} 
\hspace*{-0.4cm} \epsfig{file=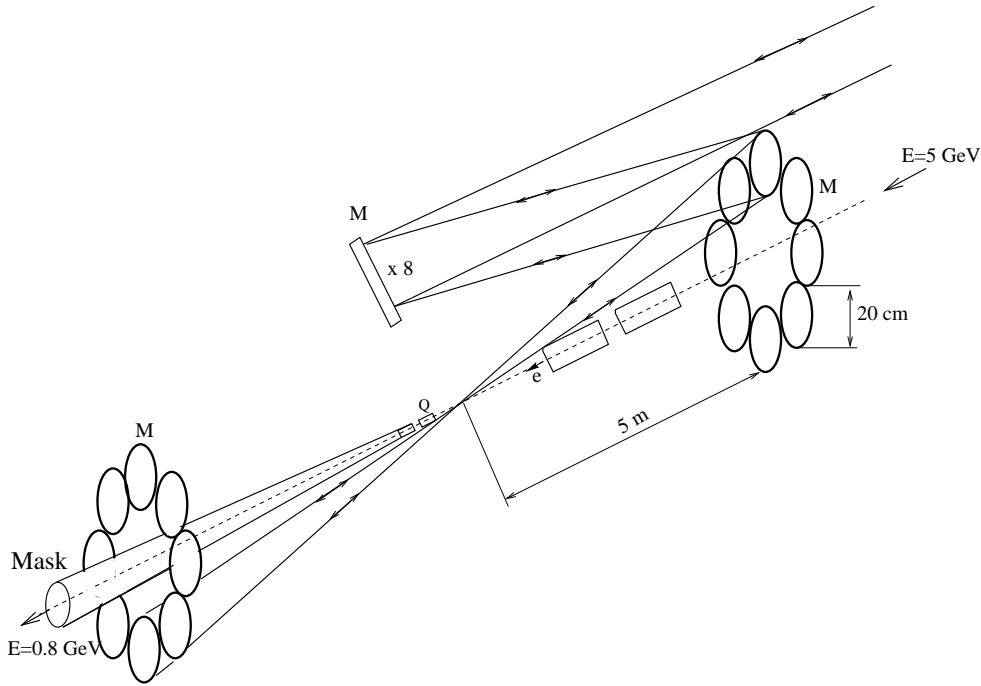,width=13cm,angle=0} 
\vspace*{-0.0cm} 
\caption{Possible scheme of the laser cooling.}
\vspace{0mm}
\label{cooling}
\vspace{4mm}
\end{figure}

The possible optical scheme for the TESLA project is shown in
fig.\ref{cooling} (only the final focusing
mirrors are shown). The system consist of 8 independent identical
optical cavities focusing the laser beams to the points distibuted
along the beam direction on the length $\Delta z \sim 2$ mm. The
length of the cavity (the distance between the ``left'' mirror and an
entrance semi-transparent mirror (not shown)) is equal to half the
distance between the electron bunches in the train, 50 m for TESLA).
The large enough angle between the edges of the mirrors and the beam
axis (30 mrad) makes X-ray flux rather small (see the estimation
above).  Also this clear angle allows the final quads to be placed at
a distance about 50 cm (from the side of the cooled beam), much closer
than the focusing mirrors. Smaller focal distance makes the problem of
chromaticity correction easier.

The maximum distance from the CP to the mirrors is determined only by
the mirror size, the diameter of 20 cm seems reasonable, which gives
$L=5$ m.  The laser spot size at the CP is 7.5 \MKM, at least 3 times
larger than the horizontal electron beam size with $\beta_x <$ 5 mm.
The circulating flash energy in each cavity is 25 J and 200 J in the
whole system, not small. The average power circulating inside the system is
$200\times 15$ kHz = 3 MW!  However, if the Q factor of the cavities
is about 1000--3000 (3000 bunches in the electron train at TESLA), the
required laser power is only 1--3 kW, or 0.15--0.4 kW/per each laser,
that is already reasonable.

What about damage to the mirrors by such powerful laser light?  The
maximum laser flash energy/cm$^2$ on the mirrors is 0.13 J/cm$^2$
(0.7-2 has been achieved for 1 ps pulses~\cite{NLC}), the average
power/cm$^2$ is 2 kW/cm$^2$ (there are systems with $>5$ kW/cm$^2$
working long time~\cite{NLC}). The average power inside one train
($\Delta t = 1$ msec) is 200 times higher (400 J/cm$^2$/1 msec), but from the
same ref.\cite{NLC} is known that 100 J/cm$^2$ for a time of 100 ns is OK,
and extrapolating as $\sqrt{t}$ (thermoconductivity) one can expect
the limit of about 10 kJ for 1 msec, much larger that expected in our
case. One of potential problems is the vatiation of the laser
amplifire temperatute inside one beam train, that is not simple to
correct by adaptive optics. Note, here we are speaking about
circulating, not absorbed energy.  So, all power densities are below
the known limits, this all depends, of course, on specific choice of
mirrors.

At last, the main numbers. After one stage of such a cooling system
the normalized emittance is decreased by a factor of 6. The ultimate
normalized emittance (after several cooling sections) is proportional
to the $\beta$-function at the CP, at $\beta_{x,y}=1$ mm it is about
$2\times10^{-9}$ m rad, smaller than can be produced by the TESLA
damping ring by a factor of 5000(15) in x(y) directions. From this
point of view such a small $\beta_x$ is not necessary, but it should be
small enough ($< 5$ mm to have a small electron spot size in the
cooling region.  The first stage of cooling will be the most efficient
because  the beam is cooled in both horizontal and vertical directions (far
from the limits). Besides, after decreasing the horizontal emittance the
$\beta$- function at the LC final focus can be made as small as
possible, $\sim \sigma_z.$ All together this can give a factor of ten in
the luminosity.

\section{Conclusion}

The laser cooling of electron beams allows to reach a very
luminosity at the high energy gamma-gamma colliders, 1--2 orders high
than it is possible without such cooling. The method is quit
straighforward, but the task is quit chelenging due to very high
required laser power (peak and average). There are hopes that this
problem can be solved using pulse stacking of laser pulses in an
``external'' optical cavity. This requires intensive R\&D.  


\begin{thebibliography}{99}
%

\bibitem{NLC} {\it Zeroth-Order Design Report for the Next Linear Collider} 
LBNL-PUB-5424, SLAC Report 474, May 1996.
\bibitem{TESLA}
{\it  Conceptual Design of a 500 GeV Electron Positron Linear Collider with 
Integrated X-Ray Laser Facility} DESY 97-048, ECFA-97-182. 
R.Brinkmann et al., {\it Nucl. Instr. {\rm\&}Meth. A}
 {\bf 406} (1998) 13.
\bibitem{JLC} {\it JLC Design Study}, KEK-REP-97-1, April 1997.
I.Watanabe et. al.,KEK Report 97-17.
\bibitem{GKST81} I.Ginzburg, G.Kotkin, V.Serbo, V.Telnov,{\it Pizma ZhETF},
{\bf 34} (1981)514; {\it JETP Lett.} {\bf 34} (1982)491.
\bibitem{GKST83} I.Ginzburg, G.Kotkin, V.Serbo, V.Telnov,{\it Nucl.Instr.
{\rm\&} Meth.} {\bf 205} (1983) 47.
%
\bibitem{GKST84} I.Ginzburg, G.Kotkin, S.Panfil, V.Serbo, V.Telnov,
 {\it Nucl.Instr.{\rm\&}Meth.} {\bf 219}(1984)5.
%
\bibitem{TEL90} V.Telnov,{\it Nucl.Instr.{\rm\&}Meth.A} {\bf 294}
(1990)72.
%
\bibitem{TEL95} V.Telnov,
 {\it Nucl.Instr.{\rm\&}Meth.A} {\bf 355}(1995)3.
%
\bibitem{BERK} {\it Proc.of Workshop on \GG\ Colliders},
Berkeley CA, USA, 1994, {\it Nucl. Instr. {\rm\&}Meth. A}
 {\bf 355}(1995).
%
\bibitem{ee97} V.Telnov, {\it Int. J. Mod. Phys.} A 13 (1998) 2399,
e-print:hep-ex/9802003.
%
\bibitem{TKEK} V.Telnov, Proc. of 17th Intern. Conference on
High Energy Accelerators (HEACC98), Dubna, Russia, 7-12 Sept. 1998,
KEK preprint 98-163, e-print: hep-ex/9810019.
%
\bibitem{Tfrei} V.Telnov, Talk at International Conference on the 
Structure and Interactions of the Photon (Photon 99), Freiburg,
Germany, 23-27 May 1999. Submitted to Nucl.Phys.Proc.Suppl.B, 
e-print: hep-ex/9908005 
%
\bibitem{Tsit1} V.Telnov, To be published in the proceedings of
  World-Wide Study of Physics and Detectors for Future Linear
  Colliders (LCWS 99), Sitges, Barcelona, Spain, 28 Apr - 5 May 1999, 
e-print: hep-ex/9910010.
%
\bibitem{TSB2} V.Telnov, {\it Proc. of ITP Workshop ``Future High energy
colliders''} Santa Barbara, USA, October 21-25, 1996,  AIP Conf. Proc. No 397, 
p.259-273; e-print: physics/ 9706003.
%
\bibitem{WID} H.Wiedemann, {\it Particle Acc. Physics: basic principles
    and linear beam dinamics}, Springer-Verlag, 1993.
%
\bibitem{TRAV} C.Travier, {\it Nucl.Instr.{\rm\&}Meth.A} {\bf 340}(1994)26.
%
\bibitem{TSB1} V.Telnov, SLAC-PUB-7337, {\it Phys.Rev.Lett.}, {\bf 78}
  (1997) 4757, erratum ibid 80 (1998) 2747, e-print: hep-ex/9610008.
%
\bibitem{Monter} V.Telnov, {\it Proc. Advanced ICFA Workshop on
Quantum aspects of beam physics,} Monterey, USA, 4-9 Jan. 1998, World
Scientific, p.173, e-print: hep-ex/9805002.
%
\bibitem{PALMER} R.Palmer {\it Nucl.Instr.{\rm\&}Meth.A} {\bf 355}(1994)150.
%
\bibitem{Tsit2} V.Telnov, To be published in the proceedings of
  World-Wide Study of Physics and Detectors for Future Linear
  Colliders (LCWS 99), Sitges, Barcelona, Spain, 28 Apr - 5 May 1999, 
e-print: hep-ex/9910011.
%
\bibitem{LAN} V.Berestetskii, E.Lifshitz and L.Pitaevskii, {\it Quantum
Electrodynamic}, Pergamont press, Oxord, 1982.
%
\bibitem{KOT} G.Kotkin, S.Polityko, V.Serbo, {\it  Nucl.Instr.{\rm\&}Meth.A}
 {\bf 405}(1998) 30.
%
\bibitem{Perry} K. van Bibber, M.Perry, Talk at Intern. Workshop on Physics and
experiments at Linear Colliders, Sitges, Spain, April 28, 1998.
%
\bibitem{Las1} V.Telnov, in Proc. of this Symposium.
\bibitem{HAAR}
  T.Smith, P.Haar, H.Schwettman, {\it Nucl. Instr. {\rm\&}Meth. A}
  {\bf 393} (1997) 245 .
\end{thebibliography}
\end{document}